\documentclass[final,3p,times,draft]{elsarticle}

\usepackage{amssymb}
\usepackage{wasysym}

\journal{Physica B}

\newcommand{\C}[1]{CeCoIn$_{5}$\,\,}

\begin{document}

\begin{frontmatter}



\title{Unconventional superconductivity in the strong-coupling limit
for the heavy fermion system \C}

\author[BT,IB]{Y.~Fasano$^{*}$}
\author[CLTP]{P. Szab\'{o}}
\author[CLTP]{J.~Ka\v{c}mar\v{c}\'{i}k}
\author[CLTP]{Z.~Pribulov\'{a}}
\author[BT,IB]{P.~Pedrazzini}
\author[CLTP]{P.~Samuely}
\author[BT,IB]{V.F.~Correa}

\address[BT]{Low Temperature Lab, CAB-CNEA, Bariloche \& CONICET, Argentina }
\address[IB]{Instituto Balseiro, UN Cuyo \& CNEA, Bariloche, Argentina}
\address[CLTP]{Centre of Low Temperatures Physics, IEP-SAS, Ko\v{s}ice, Slovakia}

\begin{abstract}

We present scanning tunneling spectroscopy measurements of the local
quasiparticles' excitation spectra of the heavy fermion \C\ between
440\,mK and 3\,K in samples with a bulk $T_{\rm c}=2.25$\,K. The
spectral shape of our low-temperature tunneling data, quite textbook
nodal-$\Delta$ conductance, allow us to confidently fit the spectra
with a d-wave density of states considering also a shortening of
quasiparticles' lifetime term $\Gamma$. The $\Delta(0)$ value
obtained from the fits yields a BCS ratio $2\Delta/kT_{\rm c} =7.73$
suggesting that \C\ is an unconventional superconductor in the
strong coupling limit. The fits also reveal that the height of
coherence peaks in \C\ is reduced with respect to a pure BCS spectra
and therefore the coupling of quasiparticles with spin excitations
should play a relevant role. The tunneling conductance shows a
depletion at energies smaller than $\Delta$ for temperatures larger
than the bulk $T_{\rm c}$, giving further support to the existence
of a pseudogap phase that in our samples span up to $T^{*}\sim 1.2
T_{\rm c}$. The phenomenological scaling of the pseudogap
temperature  observed in various families of cuprates,
$2\Delta/kT^{*} \sim 4.3 $, is not fulfilled in our measurements.
This suggests that in \C\ the strong magnetic fluctuations might
conspire to close the local superconducting gap at a smaller
pesudogap temperature-scale than in cuprates.
\end{abstract}


\end{frontmatter}
$^{*}$ corresponding author: yanina.fasano@cab.cnea.gov.ar, Low
Temperature Lab, Centro At\'omico Bariloche, Avenida Bustillo 9500,
8400 Bariloche, Argentina; +54-(0)294-4445171.

\section{Introduction}

Heavy fermion systems are interesting candidates to study the
microscopic coupling giving rise to unconventional superconductivity
in concomitance with magnetic order.~\cite{Flouquet2005} In
particular, \C\ is a paradigmatic heavy fermion in which magnetic
and superconducting orders seem to cooperate in a phase region close
to the upper critical field.~\cite{Kenzelmann2008}  The sub-meV
value of the superconducting gap in this material renders the task
of decoding specific information on the pairing mechanism from the
k-dependence of $\Delta$ a non-trivial one. Nevertheless, there is
evidence that the superconducting gap quite likely presents
nodes.~\cite{Izawa2001} Although \C\ has been one of the most
extensively studied heavy fermion superconductors, there are still
several open questions, particularly on the role played by the
strong antiferromagnetic spin fluctuations~\cite{Hu2012} for the
superconducting pairing glue, as well as the possible persistence of
superconducting fluctuations at $T>T_{\rm c}$.~\cite{Ernst2010}

In order to properly answer these questions, measurements on the
quasiparticles' excitation spectrum  with sub-meV resolution in
energy are mandatory.  This information can be accessed at the local
scale and in real space by means of low-temperature scanning
tunneling spectroscopy (STS). However, in \C\ this sort of data do
not abound.  Three previous STS studies
 claim nodal heavy fermion superconductivity in \C\ from either fitting
the quasiparticles' excitation spectrum~\cite{Ernst2010}, or
infering information on the k-space structure of
$\Delta$,~\cite{Allan2013} or detecting close to surface atomic
steps a spatial-evolution of low-energy excitations  that is typical
of nodal superconductivity.~\cite{Zhou2013}. Nevertheless,
performing multi-parameter modeling of the conductance depletion
 in order to extract spectroscopic information on the superconducting condensate from
these experimentally rounded~\cite{Ernst2010} or particle-hole
asymmetric~\cite{Allan2013,Zhou2013} STS spectra is a strategy that
can be further improved in order to produce more confident results.
In this work we measured low-temperature STS spectra of \C\ with
well defined features in the quasiparticle excitation spectra that
are the signature of nodal superconductivity, mounted on a
featureless background. The unprecedented d-wave textbook quality of
our spectra allow us to infer confident information on the
temperature evolution of $\Delta$ and the quasiparticle lifetime
shortening by fitting with a d-wave BCS model. The small discrepancy
between our spectra and these fits allow us to suggest the
consideration of a fitting model including the strong coupling to
spin fluctuations known to affect the shape of STS spectra even in
cuprate~\cite{Jenkins2009} and pnictide~\cite{Fasano2010}
superconductors.

\section{Experimental}

\begin{figure*}[bbb]
\begin{center}
\resizebox{\textwidth}{!}{\includegraphics{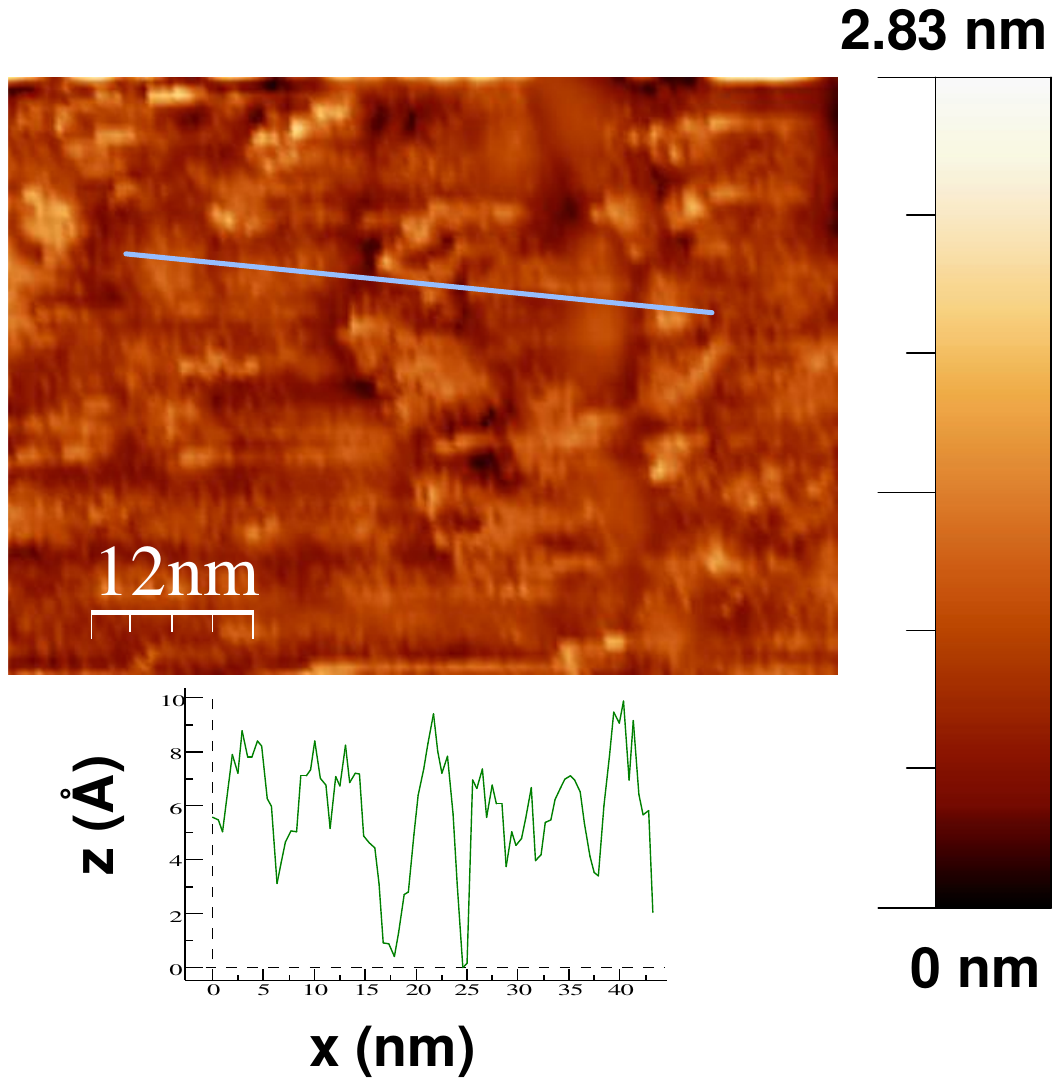}} \caption{
Typical STM topographic image of the \C\ single crystals studied.
The field of view is of $55 \times 40$\,nm$^2$. The bottom panel
shows a height profile taken along the line shown in the main
figure. } \label{fig:Figure0}
\end{center}
\end{figure*}

The local quasiparticles' excitation spectra for different
temperatures was obtained from STS measurements performed at
low-temperatures using a homemade scanning tunneling microscope
inserted in a sorption-pump based $^{3}$He fridge (Janis SSV
cryomagnetic system). This cryogenic system has a base temperature
of 300\,mK and a maximum holding time of 12\,hours. The atomically
sharp Au tip prepared in situ served as one electrode while the
sample was the ground electrode. More details on the experimental
setup can be found in Ref.\,\cite{Samuely2010}. Samples were cleaved
at room temperature with a scalpel and immediately cooled down to
low temperature. Topographic measurements in such cleaved surfaces
show atomically flat terraces of more than $30 \times 30$ nm$^{2}$
with steps between terraces of $\sim 3$ c-axis unit cells
($7.56$\,\AA) or larger, see for instance Fig.\,\ref{fig:Figure0}.
Atomic resolution was not pursued since the STM electronics was
adjusted in order to perform measuremens of average spectra in
nanometer-range fields of view as a function of temperature.

We studied several \C\ single crystals from the same batch grown
from an indium flux starting from high-purity elements as described
in Ref.\,\cite{Betancourth2015}. Platelet-like single-crystals with
the c-axis oriented perpendicular to the flat surface were obtained
after extracting the samples from the melt by using a centrifuge.
The stoichiometry and crystal structure were confirmed by means of
energy-dispersive X-ray spectroscopy and X-ray diffraction.
Energy-dispersive spectroscopy scans revealed a good homogeneity
throughout the crystals  and no traces of spurious phases were
detected. The X-ray diffraction rocking curves at different
crystallographic directions present a full-width at half-maximum of
less than 0.1\,degrees.

\begin{figure*}[bbb]
\begin{center}
\resizebox{\textwidth}{!}{\includegraphics{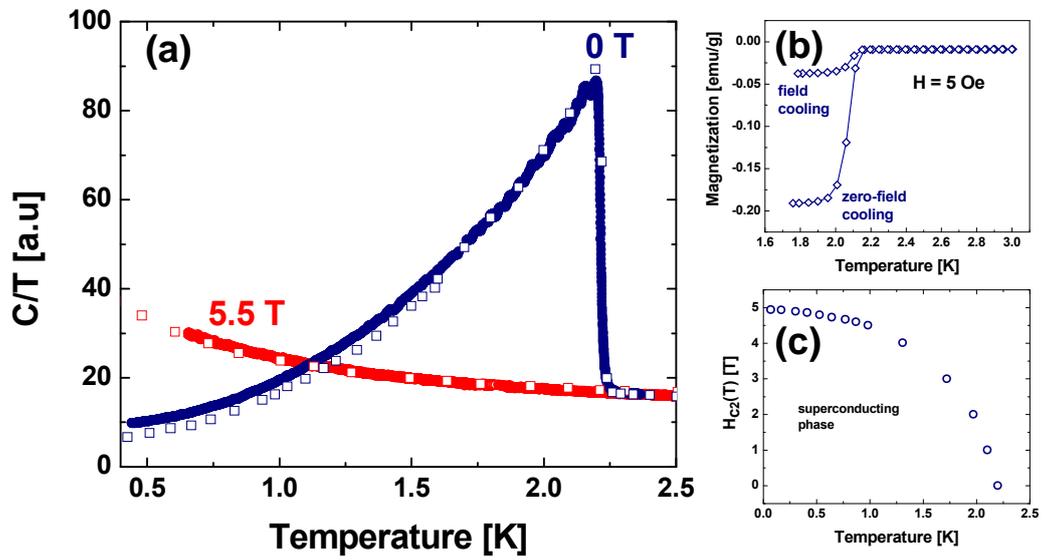}}
\caption{Characterization of the \C\ single crystals studiedby
applying bulk thermodynamic and magnetic techniques. (a)
Temperature-evolution of the ac heat capacity at 0 and 5.5\,T
(normal state data) measured at 0.4\,Hz. The transition from the
normal to the superconducting state is observed on cooling at
2.25\,K. We include dc heat capacity data published in the
literature for \C\ samples of a different sample grower for
comparison (open square points). (b) Field-cooling and zero-field
cooling branches of the magnetization of a representative crystal at
low fields. (c) Upper critical field $H_{\rm c2}(T)$ line obtained
from temperature and field-dependent ac heat capacity data as for
instance shown in (a).} \label{fig:Figure1}
\end{center}
\end{figure*}

The high-quality of the samples was further revealed by resistivity,
magnetization, thermal expansion and heat capacity measurements. The
critical temperature of the studied samples is of 2.25\,K as
detected by bulk techniques. For instance, Fig.\,\ref{fig:Figure1}
(a) shows the temperature-evolution of the ac-measured heat
capacity~\cite{Kacmarcik2016} at 0 and 5.5\,T applied along the
c-axis. At zero field, on cooling from the normal state the $C/T$
signal jumps suddenly (in less than 40\,mK) at 2.25\,K towards the
low-temperature superconducting state.  We also include for
comparison zero-field $C/T$ data from samples grown by other
group.~\cite{Petrovic2001} The field-cooling and zero-field-cooling
branches of low-field magnetization measurements split at roughly
the same transition temperature, see Fig.\,\ref{fig:Figure1} (b).
Field and temperature-dependent ac heat capacity
studies~\cite{Pribulova2017} yielded the upper critical field
$H_{\rm c2}$ shown in Fig.\,\ref{fig:Figure1} (c). These results are
in agreement with data found in the literature for other \C\
samples.~\cite{Bianchi2008}

\section{Results and discussion}

\begin{figure*}[ttt]
\begin{center}
\resizebox{\textwidth}{!}{\includegraphics{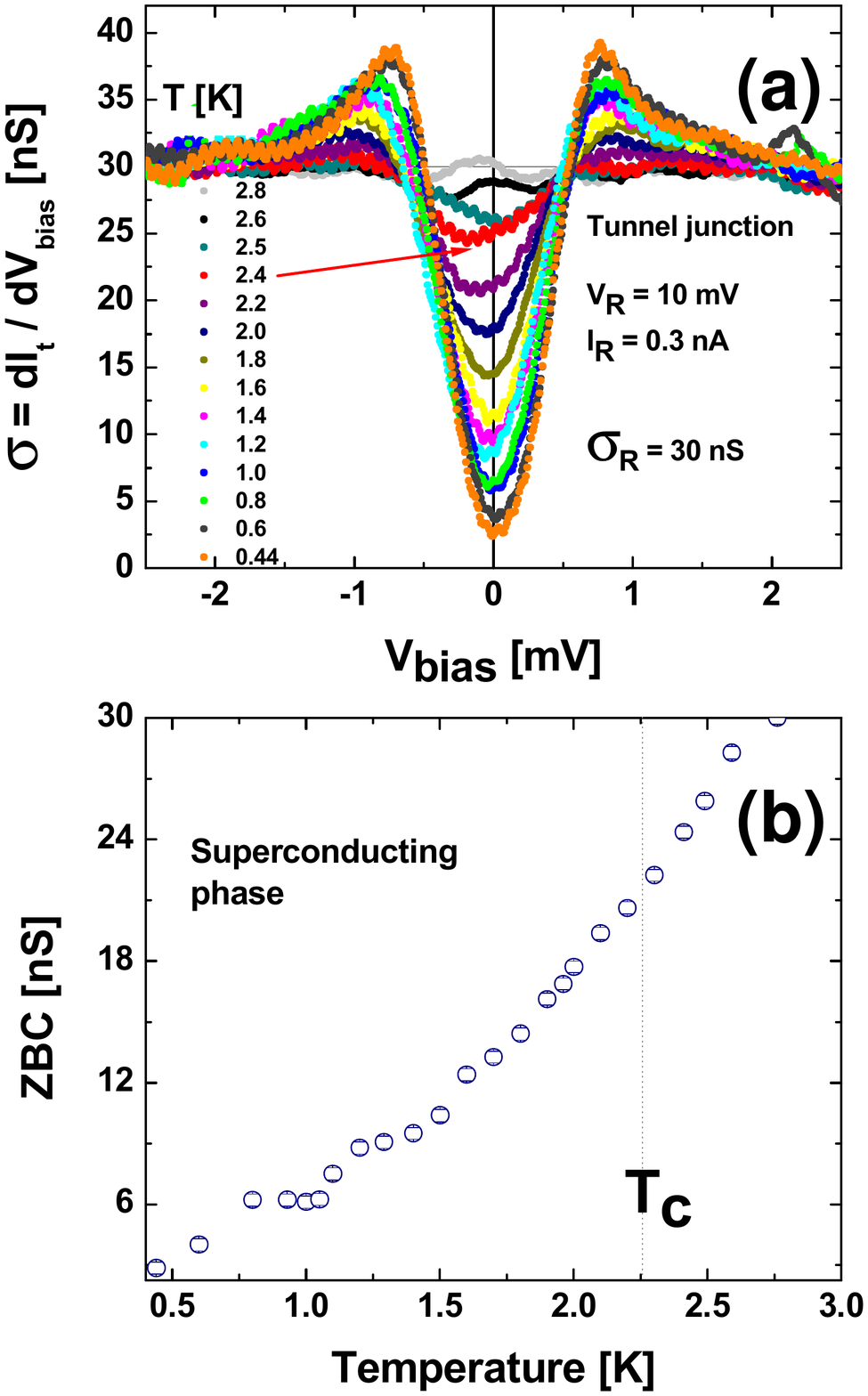}} \caption{
(a) Temperature-evolution of the STS spectra obtained from local
tunneling conductance as a function of voltage bias. The temperature
is changed between 0.2 and 1.25\,$T_{\rm c}$. The tunnel junction
regulation conditions are indicated. The arrow indicates the 2.4\,K
curve measured just above the bulk $T_{\rm c}=2.25$\,K. (b)
Evolution of the zero-bias conductance as a function of
temperature.} \label{fig:Figure2}
\end{center}
\end{figure*}

We infer the spectrum of quasiparticles' excitation from the
superconducting condensate via differential tunnel conductance  as a
function of bias voltage measurements, $\sigma(V_{\rm bias})$. We
perform experiments as a function of temperature in the range of
0.44 to 2.8\,K keeping the tunnel junction impedance constant. The
regulation tunnel current and voltage were set at $I_{\rm R} =
0.3$\,nA and  $V_{\rm R}=10$\,mV, giving a tunnel junction impedance
$\sigma_{\rm R}=30$\,nS. These parameters were chosen after
verifying that they were within the range of an exponential
dependence of the tunnel current $I_{\rm t}$ with the tip-to-sample
distance. The conductance data were obtained by deriving the
measured tunnel current as a function of $V_{\rm bias}$, namely
$\sigma = dI_{\rm t}/dV_{\rm bias}$.  Figure \ref{fig:Figure2} (a)
shows average $\sigma$ curves at different temperatures  with the
average taken over local spectrum measured in fields of view in the
nanometer range. At all temperatures, the local spectra have good
spatial reproducibility.

The low temperature $\sigma$ curves are remarkably V-shaped at low
bias and present a small zero bias conductance (ZBC). In addition,
coherence peaks are significantly developed and particle-hole
symmetric. Spectra saturate at unity at energies larger than those
of the coherence peaks when normalized by $\sigma_{\rm R}$,
indicating that the $\sigma$ signal measured at low temperatures
comes mostly from a superconducting channel of conduction and
therefore $\sigma/\sigma_{\rm R}$ is a reasonable estimation of the
quasiparticle's excitation spectrum. All these features of the
spectra are clear fingerprints of a nodal BCS superconducting
density of states.

The spectral shape of our data contrasts with the data published in
the literature in which these signatures are more elusive. For
instance, in the first report on STS data on \C\,~\cite{Ernst2010}
the tunneling conductance spectra are much more rounded, zero bias
conductance is significantly larger (roughly 10 times larger at
0.5\,K), and coherence peaks are fainted in comparison to our data.
Following works \cite{Allan2013,Zhou2013} report measurements at
$\sim 0.25$\,K and show STS spectra with rounded coherence peaks and
a zero-bias conductance of around 40\,\%. In addition, in the
$V_{\rm bias}$ range that we measure, the spectra shown in
Refs.\,\cite{Allan2013,Zhou2013} are particle-hole asymmetric and
$\sigma$ does not saturate at $V_{\rm bias} > 2$\,mV. The latter
suggests that in those works the STS spectra data might be strongly
affected by the conduction in the heavy bands.

On warming from 0.44\,K the ZBC rises at an almost linear rate,
indicating that the low-energy excitations are enhancing, see
Fig.\,\ref{fig:Figure2} (b). The coherence peaks gradually faint on
warming. Strikingly, at the bulk critical temperature the ZBC does
not saturate to the high-temperature normal state value. Indeed, a
close inspection to the spectra of Fig.\,\ref{fig:Figure2} (a)
reveals that at 2.4\,K (see red arrow), a temperature slightly
larger than the bulk $T_{\rm c}$, there is still a depletion in
$\sigma$ at low energies. This depletion persists at larger
temperatures and fills in completely at 2.8\,K within our
experimental noise. Superconducting quasiparticle excitations are
still detected at a local scale and within the same energy-scale in
a range of temperatures above $T_{\rm c}$. Therefore our data
confirm the existence of a pseudogap phase up to $T^{*}=1.2T_{\rm
c}$. The spanning of the pseudogap phase in our samples is smaller
than previously reported ($T^{*}=1.5T_{\rm c}$ in
Ref.\,\cite{Ernst2010}).

The textbook-like nodal nature of our spectra supports that fitting
the data with a d-wave BCS density of states is a quantitatively
sound analysis. We therefore fit the normalized conductance with a
model considering a superconducting $N(E)$ and a constant normal
channel $\sigma_{\rm N}$ of conductance, $\sigma/\sigma_{\rm R} =
f'(T) \cdot [N(E) + \sigma_{\rm N}]$, with $f'(T)$ the temperature
smearing term equal to the derivative of the Fermi function. The
superconducting density of states $N(E) = Re[\int^{2\pi}_{0}
d\varphi ( E - i\Gamma/2\pi\sqrt{(E - i\Gamma)^{2} - \Delta
\cos^{2}{2\varphi}})]$ takes into account the Dynes term $\Gamma$
associated to the shortening of quasiparticles' lifetime and a
d-wave $\Delta$ presenting nodes at the reciprocal space angles
$\varphi= 1, 3, 5$ and $7$ times $\pi/4$. The only input parameter
in the fits  is the measurement temperature and the output
parameters are the superconducting gap maximum $\Delta$, $\Gamma$
and a possible constant channel contribution (considered only for
formality).

\begin{figure*}[ttt]
\begin{center}
\resizebox{\textwidth}{!}{\includegraphics{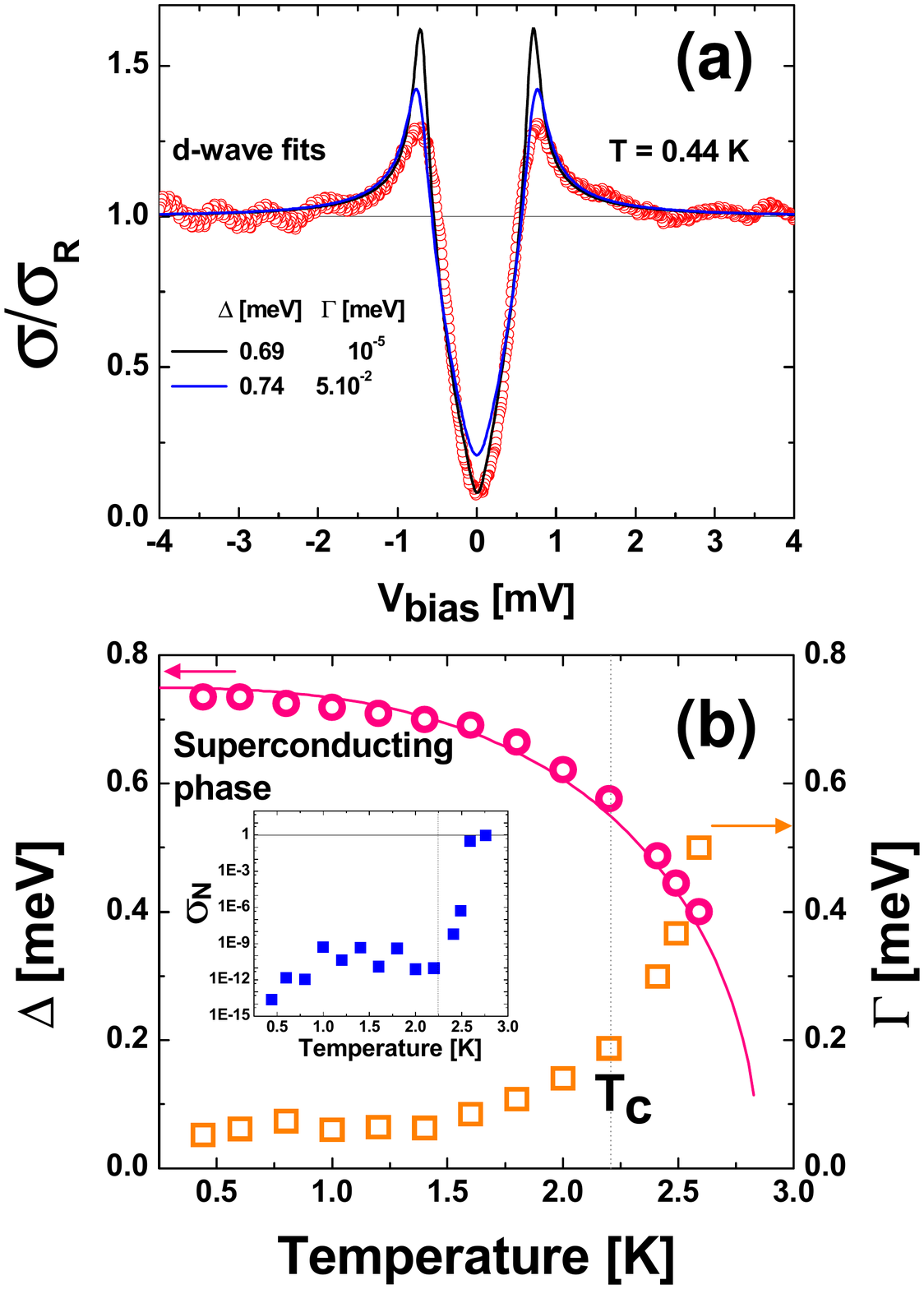}} \caption{
Fitting the quasiparticles' excitation spectrum with a d-wave BCS
density of states considering a shortening of quasiparticle's
lifetime term, $\Gamma$, and a constant channel of conduction,
$\sigma_{\rm N}$, see text for further details. (a) Spectra measured
at the lowest-studied temperature of 0.44\,K (red open circles) and
the results of fitting with two different protocols. The blue line
results from giving all experimental points the same statistical
weight whereas the black one gives 10 times more statistical weight
to the sub-gap V-shaped region of the spectra. (b)
Temperature-evolution of the superconducting gap $\Delta$ and
$\Gamma$ obtained from the blue protocol fits. The error in
determining these magnitudes is within the size of the points. The
bubble-gum-pink line is a fit to the gap data with the nodal BCS gap
temperature-evolution $\Delta(T) = \Delta(0)\sqrt{1 -
(T/T^{*})^{3}}$. Insert: temperature-evolution of the constant
normal channel.} \label{fig:Figure3}
\end{center}
\end{figure*}

Figure\,\ref{fig:Figure3} (a) shows the example of fitting the
spectra measured at 0.44\,K with two different protocols. In the
first method, the spectra are fitted considering all the measured
points in the whole  $V_{\rm bias}$ range with the same statistical
weight. This fit, shown in blue, yielded a respectable value of
quasiparticle lifetime shortening $\Gamma \sim 0.07 \Delta$ with
$\Delta = 0.735$\,meV and a negligible contribution of $\sigma_{\rm
N} = 10^{-14}$. As a consequence of minimizing the overall square
distance between all the experimental data and the model curve, the
fit significantly overestimates the ZBC and the coherence peaks'
height. The value of $\Delta$ is in a course way determined by the
energy location of the coherence peaks, whereas $\Gamma$ affects
significantly the ZBC and the height and width of the peaks.
Increasing $\Gamma$ lowers the coherence peaks but the result of the
fit will not be better: at the same time it broadens the peaks much
beyond the measured outer flanks and increases the ZBC even more. On
the other hand, decreasing the value of $\Gamma$ matches better the
low-energy V-shaped part of the spectra, but increases  too much the
height of  coherence peaks. On increasing temperature these effects
are reduced and for $T>1.6\,$K the fit matches pretty well the
experimental points even at zero bias and at the peaks.

When applying the second analysis method, the spectra are fitted
considering the data for $V_{\rm bias}$ smaller than the coherence
peaks with ten times the weight of the other points. The result of
this fitting procedure is shown in black: the low-energy excitations
part is perfectly matched by the fit at expenses of magnifying the
height of the coherence peaks. The result of this fitting protocol
yields a value of $\Gamma$ three orders of magnitude smaller and a
discrepancy with the $\Delta$ values found with the first method of
 6\% at maximum for the temperatures studied. For a given thermal
smearing the low-energy excitations are only affected by the weight
of the superconducting channel  and the value of $\Gamma$, but the
height of the coherence peaks can be modified by the coupling of the
excitations with spin collective modes detected in the material at
energies closer to $\Delta$, as well as by the normal-state
band.~\cite{Berthod2013} Therefore, for a better description and
fitting of the whole spectral shape of the quasiparticle excitation
spectra in \C\ , a fitting procedure considering a model that takes
into account the coupling with the magnetic susceptibility of the
material is required.

Nevertheless, the last approach is beyond the aim of this paper and
we will interpret the measured STS spectra by applying the first
fitting protocol (blue curve in Fig.\,\ref{fig:Figure3} (a)) that
tries to match all experimental points with the same weight. In the
worst of cases, the differences in the value of $\Delta$ yielded by
both methods is within 6\,\%. The temperature-evolution obtained
from the fits for the superconducting gap is shown in
Fig.\,\ref{fig:Figure3} (b). As expected, $\Delta$ decreases as a
function of temperature. The d-wave superconducting gap fitted at
low temperatures is smaller than some values previously reported by
STS, 0.74 against 0.9\,meV at $\sim 0.5$\,K for a sample with
$T_{\rm c}=2.3$\,K.~\cite{Ernst2010}  On the other hand, our value
is slightly larger than values reported in another STS
work~\cite{Allan2013} on samples with $T_{\rm c}=2.1$\,K. On
trespassing $T_{\rm c}$ on warming, a finite value of $\Delta$ is
still obtained from the fits up to 2.6\,K. Indeed, the depletion of
$\sigma$ at low energies disappears only at $T=2.8$\,K, see
Fig.\,\ref{fig:Figure2} (a). The temperature evolution of the gap is
reasonably well fitted by the BCS dependence for nodal
superconductors $\Delta(T) = \Delta(0)\sqrt{1 -
(T/T^{*})^{3}}$.~\cite{Dora2001} From the fit we obtain
$\Delta(0)=0.75$\,meV and a characteristic temperature for the local
closing of the superconducting gap of $T^{*}=2.85$\,K. The latter
confirms the plausible existence of a pseudogap phase but in our
samples it spans a smaller temperature range $T_{\rm c} < T <
1.2T_{\rm c}=T^{*}$, c.f. the $T^{*}=1.5T_{\rm c}$ value reported in
Ref.\,\cite{Allan2013}.

The increase of the shortening of quasiparticle's lifetime with
temperature roughly mirrors the $\Delta$ decrease, see
Fig.\,\ref{fig:Figure3} (b): $\Gamma$ slightly fluctuates around
0.075\,meV for $T<1.6$\,K and for larger temperatures presents a
steep increase reaching 0.2\,meV at the bulk $T_{\rm c}$ and even a
value of 70\,\% that of $\Delta(0)$ for $T \sim T^{*}$. This
evolution reflects the progressive rounding of the low- and
coherence peaks-energy regions of the spectra on warming. The shape
of the $\Gamma(T)$ curve also indicates that the shortening of
quasiparticles' lifetime increases dramatically on warming within
the pseudogap phase. Another magnitude that seems to be in register
with the existence of this phase is the constant normal channel of
conductance. The insert to Fig.\,\ref{fig:Figure3} (b) shows that
$\sigma_{\rm N}$ is negligible at $T<T_{\rm c}$ and therefore the
superconducting $N(E)$ term fully accounts for the measured spectra.
However, on warming beyond the superconducting phase, $\sigma_{\rm
N}$ steeply increases up to unity at 2.8\,K.

The $\Delta(0)=0.75$\,meV value obtained from the fit of the
$\Delta(T)$ curve yields a BCS ratio $2 \Delta/kT_{\rm c} = 7.73$.
Therefore our results also support that superconductivity in the
heavy fermion superconductor \C\ is in the strong coupling
limit.~\cite{Ernst2010} Nevertheless, we disagree with the
interpretation stated in Ref.\,\cite{Ernst2010} that this phase is
reminiscent to the pseudogap of underdoped cuprates. In several
families of cuprates, the pseudogap temperature follows a
phenomenological scaling $2\Delta/kT^{*}=4.3$,~\cite{Fischerreview}
suggesting that in all these materials the temperature-scale that
governs the opening of $\Delta$ at a local scale is $T^{*}$. In our
sample, the value we detect of $T^{*}$ is of 2.85\,K at best, which
gives a ratio $2\Delta/kT^{*}=6$. Therefore, a larger value of
$T^{*} \sim 4$\,K would be required in order to fulfill the
phenomenological pseudogap scaling for the BCS ratio observed in
cuprates.  This suggests that other microscopic mechanisms in
concomitance with magnetic fluctuations in this strong-coupling
limit superconductor are conspiring to locally close the gap in the
pseudogap phase at a  temperature smaller than expected for
 cuprate superconductors.

\section{Conclusions}

In summary, the unprecedent clear V-shape and small ZBC measured at
low energies, and the prominent coherent peaks detected in our
measurements of the quasiparticles' excitation spectrum of \C\ at
$T<T_{\rm c}$ provide more confidence to the interpretation of the
data via fits with a d-wave BCS density of states. We therefore
present stronger evidence on the d-wave nature of the
superconducting gap in \C\ . We confirm another previously suggested
evidence of unconventional superconductivity in \C\ , namely, the
existence of a possible pseudogap phase that in our samples would
extend up to 1.2 times the bulk $T_{\rm c}$. Since the value we
detect of $T^{*}$ is of at best 2.85\,K, this pseudogap phase does
not seem to follow the phenomenological scaling $2 \Delta/kT^{*}$
found in numerous cuprate families.~\cite{Fischerreview} Therefore
the nature of this pseudogap phase in \C\ deserves further
investigation. The value of $\Delta(0)$ provides a BCS ratio that
suggests that the material is in the strong coupling limit. The
latter, as well as the failure to perform a finer matching of the
spectra with the proposed simple model, signpost towards the need of
including the material spin susceptibility in a strong coupling
fitting analysis.~\cite{Berthod2013} This would provide valuable
information on the microscopic mechanisms governing the occurrence
of unconventional superconductivity in concomitance with magnetic
order in heavy fermion superconductors.

\section{Acknowledgements}

We acknowledge N. Haberkorn for assistance in sample preparation.
This work was supported by Argentina-Slovakia bilateral
collaboration program Conicet-SAS, by EU ERDF (European regional
development fund) Grant No. ITMS26220120047, by the Slovak Research
and Development Agency under Grant No. APVV-16-0372, by the Slovak
Scientific Grant Agency under Contract No. VEGA-0149/16 and
VEGA-0409/15, as well as by the U.S. Steel Ko\v{s}ice, s.r.o.

\end{document}